%% file: elsarticle-template-num.tex
 \documentclass[5p,times,twocolumn]{elsarticle}

\usepackage{amsmath,
            amsfonts,
            amssymb,
            latexsym,
            setspace,
            helvet}
\usepackage{extarrows}
\usepackage{amssymb}
\usepackage{epsfig}
\usepackage{graphicx}
\usepackage{epstopdf}
\usepackage{color}
\usepackage{tikz}
\usetikzlibrary{arrows,automata}
\usepackage{subfigure}
\usepackage{multirow}
\usepackage{caption}

\usepackage{listings}
\usepackage{verbatim}
\usepackage{pgfplots}

\usepackage{url}
\urldef{\mailsa}\path|{alfred.hofmann, ursula.barth, ingrid.haas, frank.holzwarth,|
\urldef{\mailsb}\path|anna.kramer, leonie.kunz, christine.reiss, nicole.sator,|
\urldef{\mailsc}\path|erika.siebert-cole, peter.strasser, lncs}@springer.com|

\newcommand{\lcolorbox}[2]{%
\hspace*{-\fboxsep}\colorbox{#1}{#2}%
}

\begin{document}
\input{macro}
\input{licsmacro}
\input{cntxmacro}

\begin{frontmatter}



\title{Automated Verification of Lock-Free Stacks by Divergence-Sensitive Bisimulation\tnoteref{label1}}
\tnotetext[label1]{
This paper is a revised version of the technical report ISCAS-SKLCS-14-16~\cite{TR14}. This research was supported  by NSFC 61100063 and by a Humboldt Fellowship (X.Y.) from Alexander von Humboldt Foundation. }


\author[labela]{Xiaoxiao Yang}
\address[labela]{State Key Laboratory of Computer Science,
            Institute of Software, Chinese Academy of Sciences}

\author[labelb]{Joost-Pieter Katoen}

\author[labelb]{Hao Wu}
\address[labelb]{Software Modeling and Verification,
           RWTH Aachen University, Germany}

\begin{abstract}
The verification of linearizability -- a key correctness criterion for
concurrent objects -- is based on trace refinement whose checking is
PSPACE-complete.  This paper suggests to use \emph{branching}
bisimulation instead.  Our approach is based on comparing an abstract specification in which object
methods are executed atomically to a real object program.
Exploiting divergence sensitivity, this also applies to
progress properties such as lock-freedom.
These results enable the use
of \emph{polynomial-time} divergence-sensitive branching bisimulation checking techniques for
verifying linearizability and progress. We conducted the experiment
on the lock-free stacks to validate the efficiency and effectiveness of our methods.

\end{abstract}

\begin{keyword}

Linearizability \sep Progress Properties \sep  Bisimulation \sep Divergence \sep Concurrency




\end{keyword}

\end{frontmatter}


\section{Introduction}

Linearizability is a standard correctness condition for the
implementation of concurrent objects \cite{Herlihy90}.
Checking linearizability amounts to verify that a concurrent object is implemented correctly by
establishing a certain correspondence between the object and its sequential specification.
In a multi-threaded environment, however, we also expect that concurrent systems satisfy liveness properties \cite{Lamport82}, which guarantee that certain "good" events can eventually happen.
This includes
progress properties \cite{Herlihy08} such as lock-freedom, wait-freedom and
obstruction-freedom, which are introduced to capture
the essence of the progress guarantee of
non-blocking concurrent algorithms \cite{Victor08}.
These algorithms employ fine-grained synchronisations instead of mutex-locks
to provide high-performance concurrent implementations of objects,
such as Michael-Scott lock-free queue, lock-free stacks \cite{hsy2004,Treiber} and Harris lock-free list
(see the \texttt{java.util.concurrent} package).
Nevertheless,
the progress property of an object
can affect the progress
of the execution of a client program that uses the object.
As Filipovi\'{e} \emph{et al.} point out
"programmers expect that the behaviour of their program
remains the same regardless of whether they use highly efficient data structures or
less-optimized but obviously correct data structures.\cite{Hearn10}"
For optimal use of concurrent objects,
it is essential that the basic properties of objects
involving linearizability and progress properties are ensured.
This emphasizes the importance of developing (automated) techniques
for verifying linearizability and progress properties.

The main work on verifying linearizability
of concurrent objects is using \emph{trace refinement} checking
\cite{Feng13,Liu13,Alur10,Hearn10,Colvin06}. This means that
a concrete implementation refines an abstract specification if and only if
the set of execution traces of the implementation is a subset of those of the specification.
Although this approach can be used well to verify linearizability,
it is not suitable for verifying progress properties.
For example, \cite{Liu13} verifies linearizability based on refinement
checking of finite-state systems specified as concurrent processes with
shared variables.
However, that refinement relation cannot be used to prove progress properties.
%
%
%
%
%
%
By the standard definition of linearizability \cite{Herlihy90},
a sequential object specification is employed as the abstract specification, to which any implementation should conform.
However, the sequential specification cannot specify liveness.
To remedy this, \cite{Gotsman11} proposes a generalisation of
linearizability that can specify liveness while \cite{Feng}
uses atomic blocks to describe the abstract specification where
each method body of an object method
should be atomic.
Although \cite{Feng,Gotsman11}
discuss progress properties of objects based on some
termination-sensitive contextual refinement,
they need to define different refinement relations
for progress properties.

Instead of the refinement relation between two objects with different granularities of an atomic operation, an
alternative perspective is to capture the relationship by means of appropriate equivalence relation.  This is justified
by the fact that client programs expect that the observable behaviour of
concrete object programs is equivalent to that of abstract ones.  The
\emph{main} issue is that whether there exists a proper equivalence relation for the object implementation,
and if there exists, what kind of equivalence is needed.
On the one side, it needs to abstract from internal steps of concrete objects as
they typically contain implementation details that are not present in
abstract objects. On the other side, we expect that such equivalence
preserves some desired properties such that for equivalent abstract and
concrete programs, the properties can be checked on the -- simpler --
abstract program rather than on the concrete program.

We are therefore motivated to explore the equivalence relation on the externally behavior of objects,
and investigate a convenient and efficient method to verify both linearizability
{\em and} progress properties of concurrent objects.
Our method employs a tailored version of branching bisimulation
\cite{Glabbeek96, Clarke88, Namjoshi97}
to analyze the behaviour of concurrent objects.
{\em Branching bisimulation} is an action-based
version of stutter bisimulation \cite{Clarke88, Namjoshi97} that
basically allows to abstract from sequences of internal steps.
This idea was first proposed in our technical report \cite{TR14}.
In this paper, we present the inspiration and motivation
of the idea in \cite{TR14}, and shows the verification method on the non-blocking stacks in detail
by directly establishing the divergence-sensitive branching bisimulation between the stack and its specification.
Our main contributions are summarized as follows:

\begin{itemize}
  \item We use the atomic block to specify an abstract object program.
        We observe that the \emph{effect} of the method call of concrete (fine-grained) objects can be abstracted into the high-level atomic block of the abstract object
        program such that the fine-grained implementation takes the same effect as the abstract atomic operation. Since the effects of method calls
        can be equivalently characterized by the observable actions call and return,
        we show that branching bisimulation is a precise notion to capture the externally
        observable equivalence between the concrete (fine-grained) object and the abstract (atomic) object.
        We further show that branching bisimulation  preserves linearizability.

  \item To verify progress properties,
        we explore the divergence-sensitive equivalence for the abstract and concrete objects.
        Divergence-sensitive branching bisimulation implies the preservation of $\mbox{LTL}_{\_\bigcirc}$ (the next-free LTL) equivalence, so it preserves various
        desirable progress properties. 

  \item Instead of trace refinement technique which is computed in PSPACE-complete, our method enables the use of
        branching bisimulation checking techniques and tools for proving correctness and progress of objects,
        which can be computed in \emph{polynomial time} for finite systems.

  \item We successfully apply branching bisimulation with explicit divergence to verify linearizability and lock-free property of
        non-blocking stacks \cite{Treiber,Michael04},
        and reveal a new bug violating the lock-freedom in the revised stack \cite{DBLP:conf/concur/FuLFSZ10}.
        Thus divergence-sensitive branching bisimulation provides us a unified framework to verify
        linearizability while preserving progress properties that are specified in $\mbox{LTL}_{\_\bigcirc}$.
\end{itemize}


\paragraph*{Overview} Section 2
defines the object system.
Section 3 introduces linearizability and the trace refinement.
Section 4 shows branching bisimulation for concurrent objects. Section 5 explores
the divergence-sensitive branching bisimulation to verify progress properties.
Section 6 conducts the experiment on verifying the non-blocking stacks.
Section 7 makes the conclusion.


\section{Preliminaries}

\subsection{Abstract and Concrete Objects}
We use two kinds of descriptions for concurrent objects:
\emph{abstract} and \emph{concrete}.
Abstract objects can be regarded as a concurrent specification,
where each method body of every object method is a single atomic operation.
Concrete objects involves more intricate interleavings that refines the implementation of abstract objects by
replacing the single atomic operation of each method body with fine-grained lock (or lock-free)
synchronization technique.
For abstract and concrete objects, each method starts with
a method call, and ends with a return action.

For instance, an abstract counter with method $\texttt{inc}()$ is given in Figure \ref{abs-con} $(i)$, where $t$ is a local variable and $c$ a shared variable,
statements embraced by \texttt{atomic}\{ \} is an atomic operation.
A concrete counter with the {\texttt{CAS}} instruction is given in Figure \ref{abs-con} $(ii)$.
The {\texttt{CAS}} instruction allows synchronization between threads at a finer grain.

\begin{figure} [ht]
\begin{minipage}{.45\linewidth}
 \vspace{0ex}
\begin{lstlisting}[language={java},  basicstyle=\small\ttfamily, keywordstyle=\color{black}, commentstyle=\color{red!50!green!50!blue!50},  rulesepcolor=\color{red!20!green!20!blue!20}]
(i) int inc()
    atomic {t:=c;
      c:=t+1;}
    return;
\end{lstlisting}
\end{minipage}
\begin{minipage}{.45\linewidth}
 \vspace{0ex}
\begin{lstlisting}[language={java},  basicstyle=\small\ttfamily, keywordstyle=\color{black}, commentstyle=\color{red!50!green!50!blue!50},  rulesepcolor=\color{red!20!green!20!blue!20}]
(ii) int inc()
     do t := c;
     while(CAS(c,t,t+1))
     return;
\end{lstlisting}
\end{minipage}
\caption{Abstract counter $(i)$ and concrete counter $(ii)$.}
\label{abs-con}
\end{figure}


\subsection{Lock-Freedom and Wait-Freedom}

Lock-free and wait-free properties are non-blocking progress conditions.
Informally, an execution of object methods is {\em lock-free} if it guarantees that some thread
can complete any started operation on the shared data structure in a finite number of steps \cite{Herlihy08}.
An execution is {\em wait-free} if it guarantees that every thread will complete an operation
in a finite number of steps \cite{Herlihy08}.

Suppose that the counter object provides a unique method $\texttt{inc}()$
for incrementing a shared variable $c$.
Figure 1 $(i)$ shows a wait-free implementation in which the increment is done atomically.
Figure 1 $(ii)$ presents a lock-free implementation by using the
{\texttt{CAS}} instruction.

\begin{Expl} \rm
Consider the following client program:

\begin{lstlisting}[language={java},  basicstyle=\footnotesize\ttfamily, keywordstyle=\color{black}, commentstyle=\color{red!50!green!50!blue!50},  rulesepcolor=\color{red!20!green!20!blue!20}]
while (1) do inc(); || inc(); print(z:=0);
\end{lstlisting}
where $||$ denotes the parallel composition.
If the client employs the wait-free counter in Figure 1 $(i)$,
then the right thread can always terminate and prints ${z:=0}$ since the wait-freedom of $\texttt{inc}()$
guarantees the termination.
But if the client uses the lock-free counter in Figure 1 $(ii)$,
the execution of $\texttt{inc}()$ by the right thread
may be non-terminating,
since the {\texttt{CAS}} instruction in Figure 1 $(ii)$ may continuously fail such that the left thread
can continuously update shared variable $c$. Thus the right thread never prints $z:=0$.
\qed
\end{Expl}

In practice, 
to achieve the desired progress properties of concurrent objects,
the garbage collection mechanism with a proper progress condition also needs to be considered.

\begin{Expl} \label{expl-2}\rm
Consider the concurrent stack with hazard pointers \cite{Michael04}, where function \texttt{Retire\_Node} reclaims the released nodes to avoid the ABA-error.
To guarantee the lock-free property of the stack, the garbage collector \texttt{Retire\_Node} needs to be wait-free.
However, if \texttt{pop} invokes another garbage collector in \cite{DBLP:conf/concur/FuLFSZ10}, which is not wait-free and not lock-free,
then it will result in the unexpected non-terminating execution of \texttt{pop} such that the implementation of the concurrent stack in \cite{DBLP:conf/concur/FuLFSZ10} violates the lock-free property. The detail is shown in Section 6.
\qed
\end{Expl}


\subsection{Object Systems}

For analysing concurrent objects,
we are interested in the implementation of objects ({\em i.e.,} object internal actions) and
the possible interactions ({\em i.e.,} call and return)
between the client and the object.
For clients, object actions are internal and usually
regarded as invisible actions, denoted by $\tau$.

Let ${\cal C}$ be a fixed collection of objects.
We have a set of actions ${Act}$ in the form of:
\vspace{-.6em}
$$
{\small Act :: = \tau \mid (t, \textsf{call}, o.m(n)) \mid (t, \textsf{ret}(n'),{o.m}) }
\vspace{-.6em}
$$
where $o \in {\cal C}$ and $t$ is the thread identifer. Silent action $\tau$ is invisible and the other two actions are visible,
where $(t, \textsf{call}, o.m(n))$ means thread $t$ invokes method call method $m(n)$ of object $o$ with parameter $n$;
and $(t, \textsf{ret}(n),{o.m})$ means thread $t$ returns the value $n'$ for  method $m$ of object $o$.

We assume there is a underlying programming language
to describe the object program. The language equipped with the operational semantics can generate a state transition system,
which is called the \emph{object system}. As in \cite{Gotsman11,Feng13}, to generate the object behaviour, we assume \emph{the most general
client} that repeatedly invokes the object's methods in any order and with all possible parameters.

\begin{definition}[Object systems] \label{lts}
Let ${\cal C}$ be a fixed collection of objects.
An \emph{object system} $\Delta$ is a labelled transition system $(S, \longrightarrow, Act, \varsigma)$,
where
\begin{itemize}
  \item[$\bullet$] $S$ is the set of states; \vspace{-.6em}
  \item[$\bullet$] $Act = \setof{\tau, ~ (t, \textsf{call}, \ $o$.m(n)), ~(t, \textsf{ret}(n'), \ $o$.{m})}{o \in {\cal C}, t \in \{1 \ldots k\}}$, where $k$ is the number of threads, $Act$ is the set of actions; \vspace{-.6em}
  \item[$\bullet$] $\longrightarrow\ \subseteq S \times Act \times S$ is the transition relation; \vspace{-.5em}
  \item[$\bullet$] $\varsigma \in S$ is the initial state.
\end{itemize}
\end{definition}
\vspace{-.2em}
Further, we have the following:
 \vspace{-.2em}
\begin{itemize}
  \item \emph{Abstract object system}, denoted by $\Theta$, is an object system of an abstract object, where the method body of each method is specified by an atomic operation $\texttt{atomic}\{ \}$.
  \vspace{-.3em}
  \item \emph{Concrete object system}, denoted by $O$, is an object system of a concrete object, where the method body of each method is specified by the low level synchronization mechanism.
  \item   {\em Linearizable specification}:
for a concrete object system $O$, we use  an abstract object system by turning each method body of $O$ into an atomic block~\cite{Feng,Liu13}, as the corresponding specification, which is called linearizable specification.

\end{itemize}

In the context, we sometimes use $\Delta$ to denote an (abstract or concrete) object system or the corresponding object program.

Each path of the object system is an execution of the object.
An {\em execution fragment} $\rho$ (also called a path fragment) is a finite or infinite alternating sequence
of states and actions starting in the initial state $s_0$
that is
$\rho {\small = \{ s_0 \alpha_0 s_1 \alpha_1 \cdots\\
\mid} ~ {\small (s_i, \alpha_i, s_{i+1}) \in ~\xlongrightarrow{} \} }$, where ${\small \alpha_i \in Act}, s_0 = \varsigma$.
A trace is a finite or infinite sequence of observable actions obtained from a path by abstracting away the states and $\tau$-transitions.
We shall write $s  \xrightarrow{a} s'$ to abbreviate $(s, a, s') \in \longrightarrow$.
A state $s'$ is reachable if
there exists a finite execution fragment such that
${\footnotesize s_0 \xlongrightarrow{\alpha_0} s_1 \xlongrightarrow{\alpha_1} \cdots}$${\footnotesize \xlongrightarrow{\alpha_n} s_n = s'}$.

\section{Linearizability}

We briefly introduce the standard linearizability definition \cite{Herlihy90}.
We use a history $H$, which is a finite sequence of actions call and return, to model an execution
of an object system. Given an object system $\Delta$,  its set of histories is denoted by ${\cal H}(\Delta)$.
Subhistory $H \mid t$ of $H$ is the subsequence of all actions in $H$ whose thread name is $t$.
The key idea behind linearizability is to compare concurrent histories to sequential histories.

A history is \emph{sequential} if (1) it starts with a method call, (2) calls and returns alternate in the history, and (3)
each return matches immediately the preceding method call.
A sequential history is \emph{legal} if it respects the sequential specification of the object.
A call is \emph{pending} if it is not followed by a matching return.
Let $complete(H)$ denote the history obtained from $H$ by deleting all pending calls if the method call has not taken effect yet
or adding the corresponding return action if the method call has taken effect.

We  use $e.call$ and $e.ret$ to denote, respectively, the invocation and
response events of an operation $e$.
An irreflexive partial order $<_H$ on the operations is defined as:
$(e, e') \in\ <_H$ if $e.ret$ precedes $e'.call$ in $H$.
Operations that are not related by $<_H$ are said to be \emph{concurrent} (or overlapping).
If $H$ is sequential then $<_{H}$ is a total order.
We first define the linearizability relation between histories.

\begin{definition}[Linearizability relation between histories]
$H \sqsubseteq_{\textsf{lin}} S$, read ``$H$ is linearizable \emph{w.r.t.} $S$'', if (1) $S$ is sequential, (2) $H|t = S|t$ for each thread $t$, and (3) $<_H~ \subseteq ~ <_{S}$.
\qed
\end{definition}

Thus $H \sqsubseteq_{\textsf{lin}} S$
if $S$ is a permutation of $H$ preserving (1) the order of actions in each thread, and (2) the non-overlapping method calls in $H$. Let $\Gamma$ denote the sequential specification and ${\cal H}(\Gamma)$ the set of all histories of $\Gamma$. For each $S \in {\cal H}(\Gamma)$,
$S$ is a legal sequential history. The linearizable object is defined as follows.

\begin{definition}[Linearizability of object systems] \label{linearizability}
An object system $\Delta$ is \emph{linearizable w.r.t.\ a sequential specification} $\Gamma$, if $\forall H_1 \in {\cal H}(\Delta).$ $\, \left( \exists S \in {\cal H}(\Gamma). \, complete(H_1) \sqsubseteq_{\textsf{lin}} S \right)$.\qed
\end{definition}




Linearizability can be casted as the trace refinement \cite{Feng13,Liu13}.
Trace refinement is a subset relationship between traces of two object systems,
an implementation 
and a linearizable specification.
Let $\mathit{trace}(\Delta)$ denote the set of all traces in $\Delta$.
Formally, for two object systems $\Delta_1$ and $\Delta_2$, we say
$\Delta_1$ refines $\Delta_2$, written as $\Delta_1 \sqsubseteq_{tr} \Delta_2$,
if and only if $~\mathit{trace}(\Delta_1) \subseteq \mathit{trace}(\Delta_2)$.

\begin{theorem} \cite{Liu13} \label{lin-refine} \rm
Let $\Delta$ be an object system and $\Theta$ the corresponding specification.
All histories of $\Delta$ are linearizable
if and only if $\Delta \sqsubseteq_{tr} \Theta$.
\end{theorem}

The above theorem shows that trace refinement exactly captures linearizability. A proof of this result can be found in \cite{Liu13}.

{\em Remark:}
Abstract objects with the high-level construct, {\em i.e.,} atomic blocks, provide us a concurrent specification \cite{Feng,Gotsman11,Liu13}. 
Using a sequential specification standard for linearizability, it is necessary to map the concurrent executions to sequential ones limiting our reasoning to these sequential executions  \cite{book08}. The permutation of concurrent executions to sequential ones is not convenient and even some highly-optimized objects, such as the collision array in the elimination stack \cite{hsy2004}, do not have the intuitive sequential sequences corresponding to them.
Concurrent specification has a more direct and natural relationship with the concurrent implementation.
It allows method calls not to terminate ({\em e.g.,} never return) and the execution of methods to overlap with the executions of others.
It adequately models that any method non-termination and overlapping execution interval in the implementation ({\em i.e.,} concrete object systems) can be reproducible in the specification ({\em i.e.,} abstract object systems).


\section{Branching Bisimulation for Concurrent Objects}
By observing the external behavior $call$ and $return$, when the
abstract object system $\Theta$ performs an operation in an atomic
step ${\small (t, \tau)}$ that takes effect for the abstract stack,
the concrete one $O$ may take several $\tau$-steps to complete the operation with the same effect as the atomic operation of $\Theta$.
This phenomenon results in that the atomic step of $\Theta$ can be
mimicked by a path with stutter steps of $O$ and a step that takes effect and these
stutter steps are exactly right the implementation details
of the concrete object, which do not change the state of the object.
Vice versa, a path fragment of concrete object system $O$
can also be mimicked by the atomic step of $\Theta$.
Such the corresponding relation between $\Theta$ and $O$ can be well characterized
by a natural equivalence relation, which is an instance of branching bisimulation \cite{Glabbeek96}
for concurrent objects. Branching bisimulation is
an action-based version of stutter bisimulation \cite{Clarke88, Namjoshi97}.

\begin{definition}\rm \label{sim}
Let $\Delta_i = (S_i, \longrightarrow_i, Act_i, \varsigma_i)~(i=1,2)$ be object systems (abstract or concrete).
A symmetric relation ${\footnotesize {\cal R}}\subseteq S_1 \times S_2$ is a {\em branching bisimulation}
for concurrent object, if for any $(s_1, s_2) \in {\cal R}$, the following holds:
{
\begin{enumerate}
  \item if ${\footnotesize s_1 \xlongrightarrow{a} s_1'}$,
  then there exists $s_2'$ such that
  ${\footnotesize s_2 \xlongrightarrow{a} s_2'}$ and ${\footnotesize (s_1', s_2')\in \cal{R}}$,
  where $a$ is a visible action ${\small (t, \textsf{call}, o.m(n))}$ or ${\small (t, \textsf{ret}(n'), o.m)}$;
  \\[-.5em]
  \item if ${\footnotesize s_1 \xlongrightarrow{\tau} s_1'}$, then (i) either ${(s_1', s_2)\in {\cal R}}$,
  (ii) or there exists a finite path fragment $s_2l_1$ $\ldots$ $l_ks_2'$
  (${\footnotesize k \geq 0}$) such that
  ${\footnotesize s_2 \xLongrightarrow{\tau} l_k}$ $\xlongrightarrow{\tau} s_2'$
  and $(s_1, l_i) \in {\cal R}$ $(1 \leq i \leq k)$ and ${\footnotesize (s_1', s_2')\in {\cal R}}$;
\end{enumerate}
where ${\footnotesize s_2 \xLongrightarrow{\tau} l_k}$ denotes
state $l_k$ is reachable from $s_2$ by performing zero or more $\tau$-transitions.
$\Delta_1$ and $\Delta_2$ are called {\em branching bisimilar}, denoted $\Delta_1 \sim_{\cal B} \Delta_2$,
if there exists a branching bisimulation ${\cal R}$ such that $(\varsigma_1, \varsigma_2) \in {\cal R}$.
\qed
}
\end{definition}
Note that the definition of $\sim_{\cal B}$ can alternatively be given by:
$$
\sim_{\cal B} = \bigcup \{ {\cal R} \mid {\cal R} \mbox{ is a branching bisimulation with} ~(\varsigma_1, \varsigma_2) \in {\cal R}\}
$$
By standard means, we can prove that $\sim_{\cal B}$ is an equivalence relation,
{\em i.e.,} $\sim_{\cal B}$ is reflexive, symmetric and transitive.

Relating to the implementation of concurrent objects, the intuitive meaning of
Definition \ref{sim} is explained as follows.
Condition (1) states that every outgoing transition of $s_1$
labelled with a visible action  must be matched by an
outgoing transition of $s_2$ with the same visible action.
Condition (2) has two cases:
(i) for each transition
$s_1 \rightarrow s_1'$ labelled with $\tau$, either $(s_1', s_2) \in {\cal R}$,
which means the step from $s_1$ to $s_1'$ is a stutter
step that does not change the state of the shared object;
(ii) or the transition is matched by a path
fragment $s_2 l_1 \ldots l_k s_2'$ such that $(s_1', s_2') \in {\cal R}$
and $s_2 l_1 l_2 \ldots l_k$
are a series of stutter steps and for each $l_i$, $(s_1, l_i) \in {\cal R}$
(a path is written as $s_0s_1\ldots s_n$ for short).
Note that stutter steps from $s_2$ to $l_k$
do not change the object to a new state, but the last step from $l_k$ to $s_2'$
is an atomic step at which point
the entire effect of the method call takes place,
which corresponds to the execution of the atomic block in the abstract object.
After the step that takes effect,
the method call may continue to take zero or more stutter steps to complete the remaining method call.
Since ${\cal R}$ is symmetric, the symmetric counterparts of cases (1)-(2)
also apply.


As we see, Definition \ref{sim} reveals an equivalence
relation $\sim_{\cal B}$ between (concurrent) abstract object $\Theta$
and concrete object $O$, which shifts
the \emph{linear-time} paradigm to a \emph{branching-time} paradigm.
The \emph{key} point is the condition (2). It is easy to understand the former case that $\Theta$ simulates $O$.
For the latter case that $O$ simulates $\Theta$,
condition 2-(ii) ($k \geq 0$) in Definition \ref{sim} is required, which
provides that for each state in $\Theta$, there can be matched
by a stutter sequence of $\tau$-steps in $O$.

\begin{theorem} \label{lin} \rm
Let $\Theta$ be an abstract object system. If $O \sim_{\cal B} \Theta$, then $O$ is linearizable.
\end{theorem}

\begin{Proof}
Because $O \sim_{\cal B} \Theta$,
from Definition \ref{sim}, it is easy to see $\sim_{\cal B}$ preserves trace equivalence, i.e.,
$trace(O) = trace(\Theta)$. Therefore, $O \sqsubseteq_{tr} \Theta$.
By Theorem \ref{lin-refine}, $O$ is linearizable.
\qed
\end{Proof}

The result 
shows that linearizability can be proven if we can establish
the branching bisimulation equivalence $\sim_{\cal B}$ between an abstract object and a concrete object.

\section{Progress Properties}

\subsection{Divergence-sensitive branching bisimulation}

Divergence means the infinite $\tau$-path of a system.
If an object system includes an infinite internal path,  {\em i.e.,}
path that only consists of stutter $\tau$-steps, then the path
will stay forever in a loop without performing
any observable action such as ${\footnotesize(t, ret(n), m)}$. We call the stutter path \emph{divergent}.
To distinguish infinite series of internal transitions from finite ones, we treat divergence-sensitive
branching bisimulation \cite{book08,Glabbeek96}.

\begin{definition} \rm \label{def-div}
Let $\Delta_i= (S_i, \longrightarrow_i, Act, \varsigma_i)$ be object systems
and ${\cal R} \subseteq S_1 \times S_2$ an equivalence relation
(such as $\sim_{\cal B}$) on the states of $\Delta_1$ and $\Delta_2$.
\begin{itemize}
  \item A state $u$ is ${\cal R}$-divergent if there exists an infinite path fragment
  $u u_1 u_2 \ldots$ such that $(u, u_i) \in {\cal R}$ for all $i >0$.
  \item  ${\cal R}$ is divergence-sensitive if for any $(u, v) \in {\cal R}$: if
  $u$ is divergent if and only if $v$ is divergent.
\end{itemize}
\end{definition}

\begin{definition} \label{div-bis} \rm
If two states $s_1$ and $s_2$ in an object system $\Delta$ is divergence-sensitive branching bisimilar,
denoted by $s_1 \sim_{\cal B}^{div} s_2$, if there exists a divergence-sensitive branching bisimulation relation ${\cal R}$
on $\Delta$ such that $(s_1, s_2) \in {\cal R}$.
\qed
\end{definition}

Two systems $\Delta_1$ and $\Delta_2$ are divergence-sensitive branching bisimilar,
if their initial states are related by $\sim_{\cal B}^{div}$ in the disjoint union of $\Delta_1$ and $\Delta_2$.

\subsection{Discussions on divergence of object systems}

The notion $\sim_{\cal B}^{div}$ is a variant of $\sim_{\cal B}$, only differing
in the treatment of divergence.
We discuss the ability of these notions $\sim_{\cal B}$, $\sim_{\cal B}^{div}$ and $\sqsubseteq_{tr}$ on identifying the divergence of object systems.

Consider the following three counters in Figure \ref{exmp:counters} associated with methods $\texttt{inc}$ and $\texttt{dec}$,
where $\texttt{inc}$ are the same, whereas $\texttt{dec}$ are implemented with distinct progress conditions.

\begin{figure} [ht]
{\small\underline{Counter} $\Delta_1$}:\\
\begin{minipage}{.48\linewidth}
 \vspace{0ex}
\begin{lstlisting}[language={java},  basicstyle=\footnotesize\ttfamily, keywordstyle=\color{black}, commentstyle=\color{red!50!green!50!blue!50},  rulesepcolor=\color{red!20!green!20!blue!20}]
int inc()
atomic {c := c+1;}
return;
\end{lstlisting}
\end{minipage}
\begin{minipage}{.48\linewidth}
 \vspace{0ex}
\begin{lstlisting}[language={java},  basicstyle=\footnotesize\ttfamily, keywordstyle=\color{black}, commentstyle=\color{red!50!green!50!blue!50},  rulesepcolor=\color{red!20!green!20!blue!20}]
int dec()
atomic {c := c-1;}
return;
\end{lstlisting}
\end{minipage}
\end{figure}
\vspace{-2em}
\begin{figure} [ht]
{\small\underline{Counter} $\Delta_2$}:\\
\begin{minipage}{.48\linewidth}
 \vspace{0ex}
\begin{lstlisting}[language={java},  basicstyle=\footnotesize\ttfamily, keywordstyle=\color{black}, commentstyle=\color{red!50!green!50!blue!50},  rulesepcolor=\color{red!20!green!20!blue!20}]
int inc()
atomic {c := c+1;}
return;
\end{lstlisting}
\end{minipage}
\begin{minipage}{.48\linewidth}
 \vspace{0ex}
\begin{lstlisting}[language={java},  basicstyle=\footnotesize\ttfamily, keywordstyle=\color{black}, commentstyle=\color{red!50!green!50!blue!50},  rulesepcolor=\color{red!20!green!20!blue!20}]
int dec()
while (1) do {skip;}
return;
\end{lstlisting}
\end{minipage}
\end{figure}
\vspace{-2em}
\begin{figure} [ht]
{\small\underline{Counter} $\Delta_3$}:\\
\begin{minipage}{.48\linewidth}
 \vspace{0ex}
\begin{lstlisting}[language={java},  basicstyle=\footnotesize\ttfamily, keywordstyle=\color{black}, commentstyle=\color{red!50!green!50!blue!50},  rulesepcolor=\color{red!20!green!20!blue!20}]
int inc()
atomic {c := c+1;}
return;
\end{lstlisting}
\end{minipage}
\begin{minipage}{.48\linewidth}
\begin{lstlisting}[language={java},  basicstyle=\footnotesize\ttfamily, keywordstyle=\color{black}, commentstyle=\color{red!50!green!50!blue!50},  rulesepcolor=\color{red!20!green!20!blue!20}]
int dec()
while (1) do {
if (c>0) then 
    c:=c-1; break;}
return;
\end{lstlisting}
\end{minipage}

\caption{Different progress properties of methods $\texttt{dec}()$.}
\label{exmp:counters}
\end{figure}

The $\texttt{dec}$ of object
$\Delta_1$ is wait-free (and also lock-free) which always guarantees each computation makes progress. The $\texttt{dec}$ of $\Delta_2$ does nothing and never makes progress.
The $\texttt{dec}$ of $\Delta_3$ is a \emph{dependent} progress condition
that makes progress only if $c>0$.
This progress condition of $\Delta_3$ is dependent on how the system schedules threads, which is different from the lock-free and wait-free properties of $\Delta_1$.

\begin{figure*}[ht]
  \centering
  \subfigure[System  $\Delta_1$.]{\label{fig:subfig:a}
    \begin{minipage}[c]{0.3\textwidth}
   \epsfig{figure=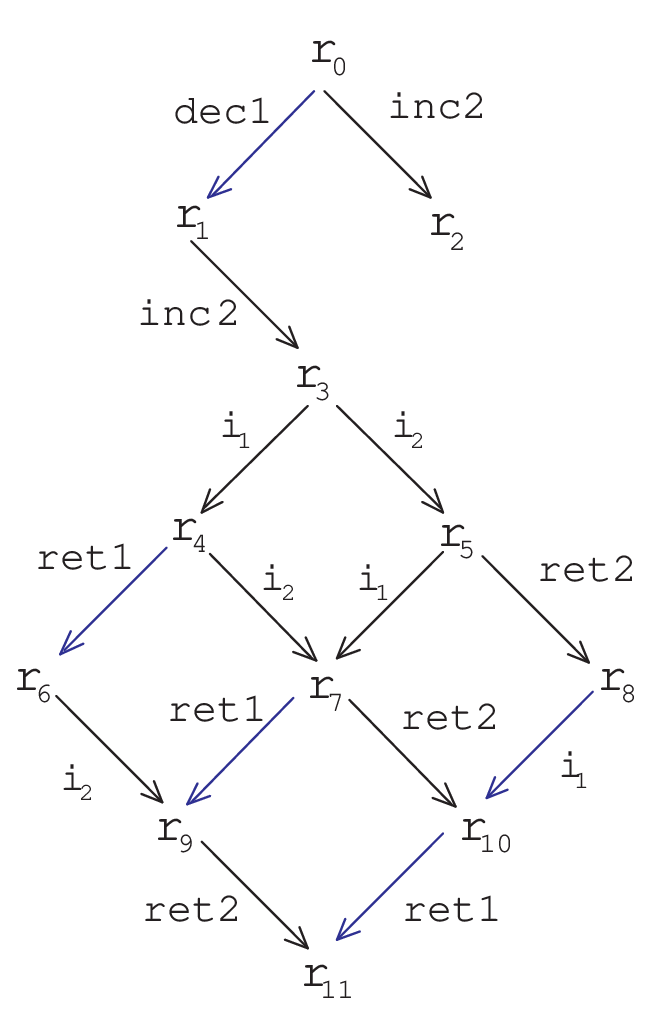,width=3.8cm}
   \end{minipage}%
   }%
   \subfigure[System  $\Delta_2$.]{\label{fig:subfig:a}
    \begin{minipage}[c]{0.3\textwidth}
   \epsfig{figure=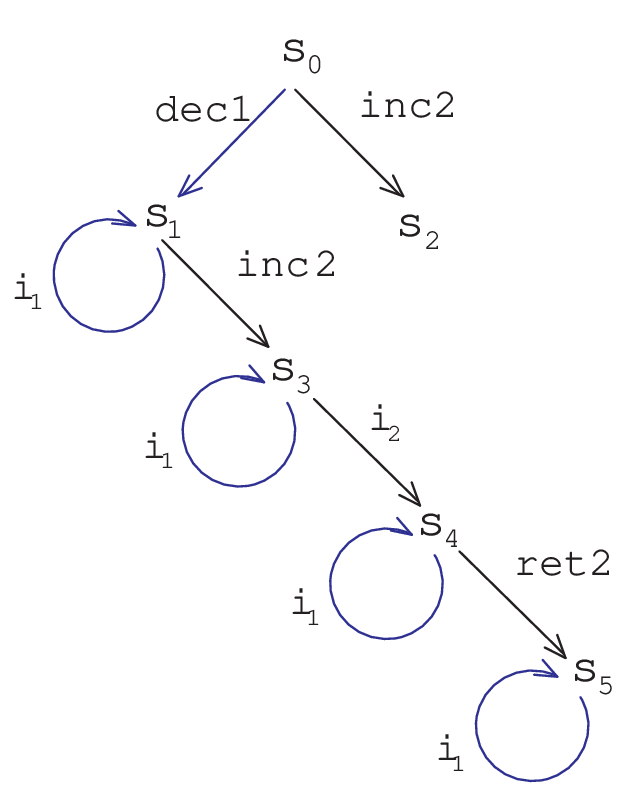,width=3.9cm}
   \end{minipage}%
   }%
   \subfigure[System  $\Delta_3$.]{\label{fig:subfig:a}
    \begin{minipage}[c]{0.3\textwidth}
   \epsfig{figure=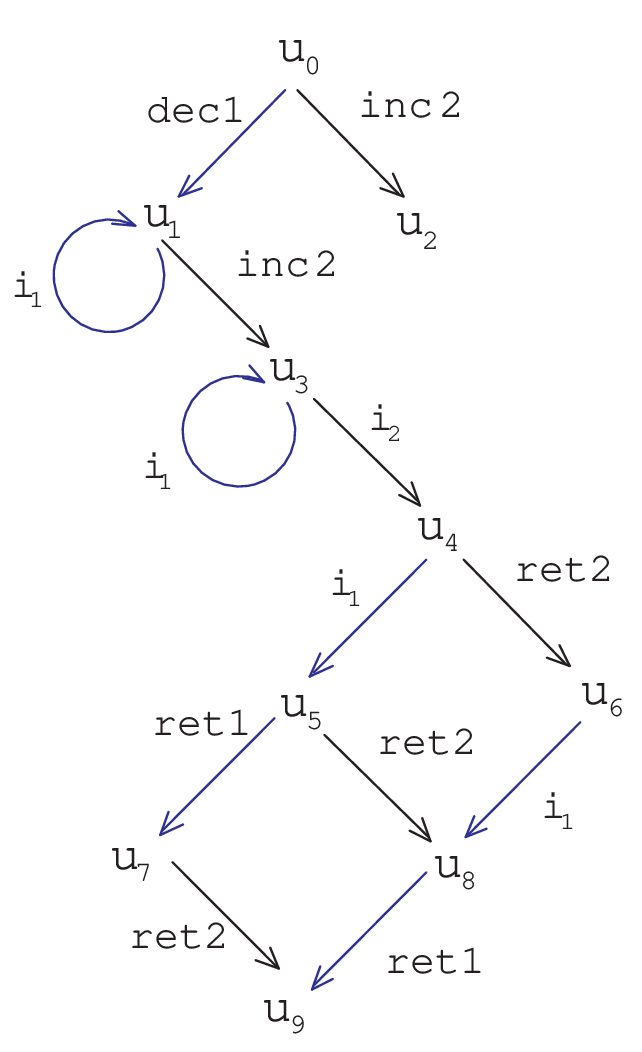,width=3.7cm}
   \end{minipage}%
   }%

   \caption{Divergence-sensitivity of bisimulation for object systems.}
  \label{diver2}
\end{figure*}

Assume that thread $t_1$ invokes $\texttt{dec}$ and $t_2$ invokes $\texttt{inc}$ once concurrently.
The transition systems for $\Delta_1$, $\Delta_2$ and $\Delta_3$ are partly presented in Figure \ref{diver2} respectively, where
we show the full transition relations from $r_3$, $s_3$ and $u_3$; and other transitions which are not relevant to our discussion are omitted.
For convenience, we write "i" to denote $\tau$, and $(t_1,call,dec)$ and $(t_1,ret,dec)$ by $dec_1$ and $ret_1$ respectively (it is the same to $inc_2$). Initially, the counter $c=0$.
Figure \ref{diver2} (a) shows the executions of $\Delta_1$, where each call by threads $t_1$ and $t_2$
can finish the execution in a finite number of steps.
In (b), the execution of $dec_1$ is a self-loop and never returns.
In (c), the execution of $dec_1$ is dependent, where at $u_1$ and $u_3$ it is a selp-loop that makes no progress,
but after $u_4$ that the atomic operation $\tau_2$ updates $c=1$,
$t_1$ evetually returns.

In  Figure \ref{diver2} (b), the self-loop of $\Delta_2$ does not generate the return action $ret_1$, whereas $\Delta_1$ in (a) can always do the return action $ret_1$. Therefore, by condition (1) of Definition \ref{sim}, $r_3 \not\sim_{\cal B} s_3$.
That is, $\sim_{\cal B} $ can distinguish the divergent state $s_3$ in (b) and non-divergent state $r_3$ in (a).

However, in Figure \ref{diver2} (c), which is a dependent progress condition,
it is easy to see $r_7 \sim_{\cal B} u_5$, so $r_3 \sim_{\cal B} u_3$.
Thus, for this case, $\sim_{\cal B}$ is not sufficient to distinguish the self-loop of state $u_3$ in $(c)$ from the state $r_3$ in (a).
However the notion of divergence-sensitive branching bisimulation $\sim_{\cal B}^{div}$ can distinguish any divergent state from non-divergent state,
that is, $u_3 \not\sim_{\cal B}^{div} r_3$.

For the refinement relation $\sqsubseteq_{tr}$, we found that it cannot distinguish any divergent states, even when the simplest case in Figure \ref{diver2} (a) and (b),
where $trace(s_0) \sqsubseteq_{tr} trace(r_0)$. Therefore relation $\sqsubseteq_{tr}$ is not suitable for checking any progress properties.

\subsection{Checking progress properties via bisimulation}

There are various progress properties of concurrent objects, such as
lock-freedom and wait-freedom for non-blocking concurrency, and
starvation-freedom and deadlock-freedom for blocking concurrency.
These progress properties can all be specified in LTL$_{\_\bigcirc}$
(LTL without next) \cite{Petrank09,icfem06}.

Because the linearizable specification $\Theta$ should allow all possible interleavings
and be able to terminate the execution of its atomic block
at any state,  there always exists some thread that can make progress and perform
the return action in any execution of $\Theta$.
This implies that any execution of abstract object system $\Theta$ is lock-free,
{\em i.e.,} any abstract object system $\Theta$ is lock-free.
Therefore, obtaining a lock-free abstract model is straightforward directly from the concurrent specification.

\begin{lemma}  \label{lemma}
The linearizable specification $\Theta$ is lock-free.
\end{lemma}

\begin{Proof}
$\Theta$ consists of a single atomic block (see Section 3.2), of which the internal execution
corresponds to the computation of the sequential specification that by assumption is always
terminating. Hence for any run of $\Theta$, there always exists one thread to complete
its method call in finite number of steps.
\qed
\end{Proof}

To obtain wait-free abstract object systems, we need to enforce some fairness constraints
to the transition systems to rule out the undesirable paths.
The common fair properties (such as strong fairness) can be expressed
in next-free LTL \cite{book08}.

It is known that divergence-sensitive branching bisimulation implies (next-free) LTL$_{\_\bigcirc}$-equivalence
\cite{book08}. This also holds for countably infinite transition systems that are finitely branching.
Thus, $O \sim_{\cal B}^{div} \Theta$ implies the preservation of all next-free LTL formulas. Since
the lock-freedom (and other progress properties \cite{Petrank09,icfem06}) can be formulated in next-free LTL, for
abstract object $\Theta$ and concrete object $O$, the lock-free property can be preserved by the relation $O \sim_{\cal B}^{div} \Theta$.

\begin{theorem} \label{lock-wait}\rm
Let $O$ be a concrete object system, and $\Theta$ the linearizable specification of $O$.
\begin{enumerate}
\item If $O \sim_{\cal B}^{div} \Theta$, then $O$ is linearizable and lock-free.
\item If $O \sim_{\cal B}^{div} \Theta$ and $\Theta$ is wait-free, then $O$ is linearizable and wait-free.
\end{enumerate}
\end{theorem}

\begin{Proof}
Because $O \sim_{\cal B}^{div} \Theta$,
from Definitions \ref{def-div} and \ref{div-bis}, it is easy to see $\sim_{\cal B}^{div}$ preserves trace equivalence, i.e.,
$trace(O) = trace(\Theta)$. Therefore, $O \sqsubseteq_{tr} \Theta$.
By Theorem \ref{lin-refine}, $O$ is linearizable. Further, because divergence-sensitive
branching bisimulation implies next-free LTL formula \cite{book08},
the lock-free property can be preserved by $\sim_{\cal B}^{div}$.
By Lemma \ref{lemma}, $\Theta$ is a lock-free abstract object, so if  $O \sim_{\cal B}^{div} \Theta$, then $O$ is lock-free.
The case of wait-free property can be proved similarly.
\qed
\end{Proof}

\begin{figure*}[ht]
\centering \scriptsize{
$$
\begin{array}{ll}
  \texttt{class Node \{ }
  \texttt{int value;}
  \texttt{~~Node next;}
  \texttt{~~Node (int value)~\{ }
  \texttt{value = value;}
  \texttt{~~next = null;}
  \texttt{~\} }
  \texttt{\} } \quad
  \texttt{Node Top;}  \quad // ~\texttt{Top} \mbox { is a shared variable. }    \\
  \texttt{void init() \{ Top := {null} \} }
  \end{array}
  $$
  }
\begin{minipage}{.24\textwidth}
\vspace{-3ex}
\begin{lstlisting}[escapechar=|, language={java}, basicstyle=\scriptsize\ttfamily, commentstyle=\color{red!    50!green!50!blue!50},  rulesepcolor=\color{red!20!green!20!blue!20}]
01 void push(v) {
02 bool done:=false;
03 Node x:=new_node(v);
04 while(!done) {
05  Node old:=Top;
06  x.next:=old;
07  |\setlength{\fboxsep}{1pt}\lcolorbox{yellow}{done:=CAS(\&Top,old,x);}|
08  }
09 return;
10 }
\end{lstlisting}
\end{minipage}
\begin{minipage}{.24\textwidth}
\begin{lstlisting}[escapechar=|, language={java}, basicstyle=\scriptsize\ttfamily, commentstyle=\color{red!50!green!50!blue!50},  rulesepcolor=\color{red!20!green!20!blue!20}]
11 int pop() {
12 bool done:=false;
13 while(!done) {
14 |\setlength{\fboxsep}{1pt}\lcolorbox{yellow}{Node old:=Top;}|
15  if (old==null)
16    then return EMPTY
17  Node x:=old.next;
18  |\setlength{\fboxsep}{1pt}\lcolorbox{yellow}{done:=CAS(\&Top,old,x);}|
19  }
20 return old.data;
21 }
\end{lstlisting}
\end{minipage}
\begin{minipage}{.23\textwidth}
\begin{lstlisting}[escapechar=|, language={java}, basicstyle=\scriptsize\ttfamily, commentstyle=\color{red!50!green!50!blue!50},  rulesepcolor=\color{red!20!green!20!blue!20}]
22 void push_HP(v) {
23 bool done:=false;
24 Node x:=new_node(v);
25 while(!done) {
26  Node old:=Top;
27  x.next:=old;
28  |\setlength{\fboxsep}{1pt}\lcolorbox{yellow}{done:=CAS(\&Top,old,x);}|
29  }
30 return;
31 }
\end{lstlisting}
\end{minipage}
\begin{minipage}{.25\textwidth}
\vspace{-2.2ex}
\lstset{escapeinside={<@}{@>}}
\begin{lstlisting}[escapechar=|, language={java},  basicstyle=\scriptsize\ttfamily, commentstyle=\color{red!50!green!50!blue!50},  rulesepcolor=\color{red!20!green!20!blue!20}]
32 pop_HP() {
33 done:=false;
34 while(!done) {
35  Node old:=Top;
36  if(old=null) return EMPTY;
37  |\setlength{\fboxsep}{1pt}\lcolorbox{yellow}{*hp:=old;}|
38  if(Top==old)
39    x:=old.next;
40    done:=cas(&Top,old,x);
41    }
42 v:=o.value;
43 |\setlength{\fboxsep}{1pt}\lcolorbox{yellow}{Retire\_Node(o);}|
44 return v;
45 }
\end{lstlisting}
\end{minipage}
\vspace{-2em}
\caption{Treiber's lock-free stack push/pop and the lock-free stack with hazard pointers push\_HP/pop\_HP, where the implementation of Retire\_Node(o) can be found in \cite{Michael04}.} \label{treiber}
\end{figure*}

\section{Experiments on Verifying Lock-Free Stacks}

To show the efficiency and effectiveness of our method for proving linearizability and progress properties,
we conduct the experiment on verifying non-blocking stacks \cite{Treiber,Michael04,DBLP:conf/concur/FuLFSZ10}.
We employ the Construction and Analysis of
Distributed Processes (CADP) toolbox \cite{cadp13} to model and verify the algorithms.

\subsection{Analyzing the concurrent stack with hazard pointers}

\paragraph*{Treiber stack}

Figure \ref{treiber} (Lines 1-21) shows the Treiber's stack algorithm \cite{Treiber},
which involves two methods $\texttt{push}$ and $\texttt{pop}$.
The stack is implemented as a linked-list pointed by the $\texttt{Top}$ pointer (shared variable).
Both operations modify the stack by doing a \texttt{CAS}.
The linearization point of $\texttt{push}$ is Line 7 if the \texttt{CAS} succeeds,
and the linearization point of $\texttt{pop}$ is either Line 14 if the stack is empty;
or Line 18 if the \texttt{CAS} succeeds.

\paragraph*{Memory reclamation - hazard pointers}

We further consider a complicate concurrent stack that involves with hazard pointers,
which provides a memory reclamation mechanism to avoid the ABA problem~\cite{Michael04}.
The hazard pointer variation is shown in Figure \ref{treiber} (Lines 22-45), which includes
$\texttt{push\_HP}$ and $\texttt{pop\_HP}$. Method $\texttt{push\_HP}$ is the same as $\texttt{push}$. Method $\texttt{pop\_HP}$ involves a hazard pointer before the ABA-prone code (Line 37), and calls the method $\texttt{Retire}\_\texttt{Node(o)}$ (Line 43) after the successful $\texttt{CAS}$ operation to determine which retired locations can be freed.
The codes of $\texttt{Retire}\_\texttt{Node(o)}$ can be found in \cite{Michael04}.
These operations are used to reclaim memory blocks safely, which is not relevant to the effect of the method call.
Further to achieve the desired lock-freedom of the stack, it is required that method $\texttt{Retire}\_\texttt{Node(o)}$ should be wait-free.

Figure \ref{spec} presents an abstract stack as the specification, where
each method body of methods $\texttt{push\_spec}$ and $\texttt{pop\_spec}$ is implemented by an atomic block.

The linearization point of $\texttt{push}$/$\texttt{push\_HP}$ and $\texttt{pop}$/$\texttt{pop\_HP}$ of the concurrent stacks is independent of the dynamic execution, which can be simply located on the implementation code such that the effect of the method call always takes place at the static linearization point between the method call and the corresponding method return. Therefore, the \emph{effect} of each mehtod call can be abstracted into the atomic operation of the linearizable specification; and vice versa, the atomic operation of the specification can be refined to several $\tau$-steps including a linearization point of the concurrent stack. Such the corresponding relation can be naturally characterized by the branching bisimulation equivalence defined in Definition \ref{sim}.
Our experiment confirms the existence of the branching bisimulation equivalence relation between the concurrent stack
(with hazard pointers) in Figure \ref{treiber} and the abstract stack in Figure \ref{spec}. To check the progress condition, the explicit divergence in branching bisimulation
is also need to be considered.

\begin{figure}[ht]
{\scriptsize
\begin{minipage}{.5\linewidth}
 \vspace{0ex}
\begin{lstlisting}[escapechar=|, language={java},  basicstyle=\scriptsize\ttfamily, keywordstyle=\color{black}, commentstyle=\color{red!50!green!50!blue!50},  rulesepcolor=\color{red!20!green!20!blue!20}]
void push_spec(int x)
|\setlength{\fboxsep}{1pt}\lcolorbox{yellow}{atomic}| {
Node node:=new Node(x);
node.next:=Top;
Top:=node;
}
return;
\end{lstlisting}
\vspace{-.2em}
\end{minipage}
\begin{minipage}{.5\linewidth}
 \vspace{0ex}
\begin{lstlisting}[escapechar=|, language={java},  basicstyle=\scriptsize\ttfamily, keywordstyle=\color{black}, commentstyle=\color{red!50!green!50!blue!50},  rulesepcolor=\color{red!20!green!20!blue!20}]
int pop_spec()
|\setlength{\fboxsep}{1pt}\lcolorbox{yellow}{atomic}| {
if (Top == null) b:=false
else
Node Curr_Top:=Top;
Top:=Curr_Top.next; b:=true
}
if (b==false)
then return EMPTY
else return Curr_Top.data
\end{lstlisting}
\end{minipage}
\vspace{-.7em}
\caption{The linearizable specification of concurrent stacks.} \label{spec}
}
\end{figure}

\subsection{Experimental results}

\begin{table*} [ht]
{\scriptsize
\begin{tabular}{|c|c|c|c|c|c|c|c|c|}

\hline \hline

\multirow{2}{*}{\#th/\#op} &

\multirow{2}{*}{$\Delta_{spec}$ } &

\multirow{2}{*}{$\Delta_{Tr}$} &

\multicolumn{2}{|c|}{Verification time (in s) ($\Delta_{Tr}$) } &

\multirow{2}{*}{$\Delta_{HP}$} &

\multicolumn{2}{|c|}{Verification time (in s) ($\Delta_{HP}$)}  \\

\cline{4-5} \cline{7-8}   &&& Thm \ref{lock-wait} (linearizability/lock-free) & Thm \ref{lin-refine} (linearizability)
&& Thm \ref{lock-wait} (linearizability/lock-free) & Thm \ref{lin-refine} (linearizability)
 \\
\hline
2/1   & 70  &  81  & 0.73  &  0.80  &  195    &  0.76  & 0.97 \\
\hline
2/2   & 487   & 823    & 0.74  & 0.92   & 7493    & 0.80   & 2.93 \\
\hline
2/3   & 1678  & 3673   & 0.81  & 2.15   & 93352   & 1.45   & 25.30  \\
\hline
2/4   & 4237  & 10999  & 0.93  & 5.08   & 808079  & 6.67   & 232.02 \\
\hline
2/5   & 8920  & 26101  & 1.04  & 11.04  & 5447816 &  101.11  & 10514.94  \\
\hline
2/6   & 16651 & 53197  & 1.20  & 21.89  & 31953747 &  283.87 & $>$16h \\
\hline
2/7   & 28516 & 97435  & 1.63  & 39.55  & 174455921 & 1528.88 & $>$10h \\
\hline
2/8   & 45769 & 164881 & 2.14  & 63.14  & M. O. & - & - \\
\hline
3/1   & 706  & 1036  &  2.25  &  2.92  & 8988  & 2.51 &  9.11 \\
\hline
3/2   & 12341 & 40309 & 1.09  & 21.48  & 4937828 & 76.26 & $>$10h \\
\hline
3/3   & 77850  &  411772 & 4.11  & 230.60  & M. O.  &  -  &  - \\
\hline
3/4   & 314392  & 2247817  & 19.90  & $>$10h  &  M. O. &  -   & - \\
\hline
4/1 &  6761 & 15595  &  0.71 & 11.36 & 612665 & 9.36 & 807.35 \\
\hline
4/2 & 304197  & 2351919  & 22.52  & 4665.74  & M. O.  &  -  & - \\
\hline
5/1 & 64351  & 261527  &  10.48  &  816 & 60598453  &  408.27  &$>$10h \\
\hline
6/1 &  616838 & 4771785  & 141.86  &  81316.58  & M. O.   &  -  & - \\
\hline
\end{tabular}
}
\vspace{-.5em}
\caption{Experimental results on verifying the lock-free stacks with/without hazard pointers.}
\label{stack-results}
\end{table*}

To verify linearizability and lock-free property of concurrent stacks,
we check that the abstract object in Figure \ref{spec} and the two concrete objects in Figure \ref{treiber} are divergence-sensitive branching bisimilar.
For the automated verification, we conduct the experiment on finite systems.
All experiments run on a server which is equipped with a 4 $\times$ 24-core Intel
CPU @ 2.3 GHz and 1024 GB memory under 64-bit Centos 7.2.

The verification results of both versions of the concurrent stacks are shown in
Table~\ref{stack-results}, where M.O.\ means memory out.
The first column indicates the number of threads and operations.
The second and third columns indicate the state space of the specification $\Delta_{spec}$ and Treiber's stack $\Delta_{Tr}$ respectively; the fourth and the fifth columns indicates the verification time (in seconds) of $\Delta_{Tr}$ based on bisimulation technique (Theorem \ref{lock-wait}) and trace refinement checking (Theorem \ref{lin-refine}). The next column shows the state space of the variant stack with hazard pointers $\Delta_{HP}$.
And the last two columns indicate the verification time (in seconds) of $\Delta_{HP}$ based
on the two kinds of methods.

From the experimental results, we can see that,
the branching bisimulation method is much more feasible and efficient than trace refinement checking for finite-state systems.
For example, based on the bisimulation equivalence checking,
the verification of $\Delta_{HP}$  with about 5 million states (in the case 2/5) takes around 100 seconds, while
trace refinement checking takes around 3 hours. For the most million states, the time of trace refinement checking is greater than 10 hours.

\vspace{-.5em}

\subsection{Bug hunting - the stack with revised hazard pointers~\cite{DBLP:conf/concur/FuLFSZ10}}
\begin{figure}[ht]
\vspace{-.5em}
\includegraphics[scale=.5]{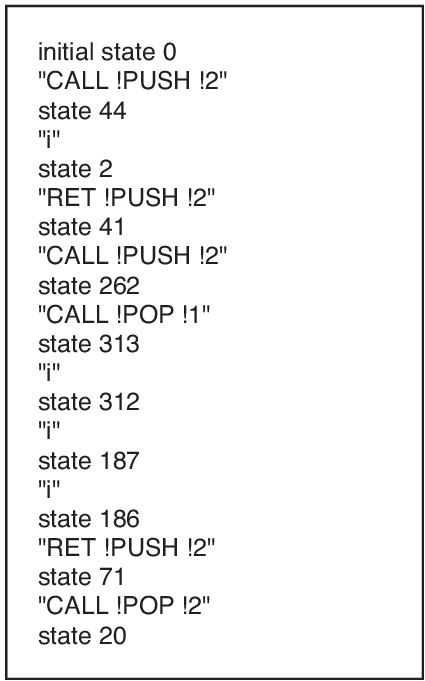}
\caption{The counter-example of trace generated by CADP.}\label{fig:tr2}
\end{figure}

While we verify the correctness of the concurrent stack with the revised hazard pointers in~\cite{DBLP:conf/concur/FuLFSZ10}, a counter-example (a trace which witnesses the concrete object and the specification are not divergence-sensitive branching bisimilar) is given by CADP with just two concurrent threads.
The revised version of the hazard pointers shown in Figure \ref{violation-lockfreedom}
avoids the ABA problem at the expense of violating the wait-free property of hazard pointers in the original algorithm \cite{Michael04}.
The error-path ends in a self-loop (at state 20 in Figure~\ref{fig:tr2}) in which one thread keeps reading the same hazard pointer value of another
thread in function $\texttt{retireNode(t)}$ without making any progress and causes the divergent path of the stack.
So the revised stack in \cite{DBLP:conf/concur/FuLFSZ10} is not lock-free.

\begin{figure}[ht]
\begin{minipage}{.23\textwidth}
\vspace{-2.2ex}
\lstset{escapeinside={<@}{@>}}
\begin{lstlisting}[escapechar=|, language={java},  basicstyle=\scriptsize\ttfamily, commentstyle=\color{red!50!green!50!blue!50},  rulesepcolor=\color{red!20!green!20!blue!20}]
pop() {
local done,next,t,t1;
done:=false;
while(!done) {
t:=Top;
if(t==null)
  return EMPTY;
|\setlength{\fboxsep}{1pt}\lcolorbox{yellow}{HP[tid]:=t;}|
t1:=Top;
if(t==t1) {
 next:=t.Next;
 done:=cas(&Top,t,next) }
}
retireNode(t);
HP[tid]:=null;
return t;
}
\end{lstlisting}
\end{minipage}
\hfill
\begin{minipage}{.2\textwidth}
\lstset{escapeinside={<@}{@>}}
\begin{lstlisting}[escapechar=|, language={java},  basicstyle=\scriptsize\ttfamily, commentstyle=\color{red!50!green!50!blue!50},  rulesepcolor=\color{red!20!green!20!blue!20}, showstringspaces= false]
|\setlength{\fboxsep}{1pt}\lcolorbox{yellow}{retireNode(t);}|
local i,t';
i:=1;
while(i<=th_num) {
if(i!=tid) {
  t':=HP[i];
  if (t'!=t) {
   i:=i+1;}
  }
else i:=1+1;
}
}
\end{lstlisting}
\end{minipage}
\vspace{-1em}
\caption{A buggy implementation of the lock-free stack \cite{DBLP:conf/concur/FuLFSZ10}.}
\label{violation-lockfreedom}
\end{figure}

\paragraph{Summary} Our experiment presents the following advantages: (i) We provide a \emph{uniform} notion $\sim_{\cal B}^{div}$ to verify both the linearizability and lock-freedom of the variant of concurrent stacks; (ii) Our techniques are fully automated (for finite-state systems) and rely on the efficient existing branching bisimulation algorithms.
(iii) In contrast to proof techniques \cite{Victor08,Feng13,Feng14} for linearizabilty and progress,
our method is able to generate counterexamples in an automated manner.


Our method is built on the behaviour of objects, not
on their statically syntactic constructs. Thus, for any
concurrent objects, we do not care how the object algorithm is
constructed, but only concerns state transition relations between the abstract and concrete
system models. Therefore, our method can be applied to general concurrent object programs.

Finally, we like to point out that divergence-sensitive branching bisimulation
implying LTL$_{\_\bigcirc}$ equivalence applies to any countable
transition system. This implies that all progress properties of
abstract object programs are carried over to bisimilar concrete ones.
For analysing infinite systems, branching bisimulation can be done using standard proof techniques such as parameterised boolean
equation systems \cite{concur07}.
For finite-state systems, polynomial-time
algorithms can be employed \cite{Groote90}. Note that in the latter setting,
trace refinement is PSPACE-complete.

\section{Related Work and Conclusion}

As we mentioned in the introduction,
almost all the work in the literature on the verification of the correctness of concurrent objects are based on the refinement techniques.
Our work takes the first step to explore the divergence-sensitive bisimulation equivalence technique
to verify linearizability and progress properties of concurrent objects \footnote{This paper is a revised version of our technical report \cite{TR14}.}.

Our transition system models are closely related to \cite{Liu13}.
The key idea of \cite{Liu13} is to construct a linearizable
specification using labeled transition systems (LTSs) and
describe linearizability as the trace refinement relation between
the specification and the concurrent algorithm. They also use
abstract objects and state transition systems as semantic models for
describing the behaviour of concurrent objects, which correspond
to our abstract and concrete object systems.
However, the trace refinement relation
they proposed is only suitable for checking linearizability, not for
checking other properties e.g., progress properties.
Moreover, they do not discuss the (bi)-simulation relations between an
abstract program and a concrete object program.

In our work, we reveal that there exists a natural (branching) bisimulation
equivalence between the abstract and the concrete object if we use a coarse-grained concurrent object as the specification.
Moreover, when do model checking,
for finite-state systems, polynomial-time branching bisimulation
checking algorithms is provided \cite{Groote90} - in contrast to (classical)
PSPACE-complete trace refinement - for checking linearizability.

Different from \cite{Liu13}, where the trace refinement is shown to be equivalent to linearizability,
our work (Theorem \ref{lin}, Theorem \ref{lock-wait}) shows that (divergence-sensitive) branching bisimulation implies linearizability.
If an object system and the specification are (divergence-sensitive) branching bisimilar,
then the object is linearizable; but if they are not (divergence-sensitive) branching bisimilar,
then it will generate a counterexample, which needs to be analyzed to determine whether it violates the linearizability (e.g., it may violate the lock-free property).

Other model checking methods on verifying linearizability are \cite{Colvin06,spin09,pldi10}.
Cerny et al. \cite{Colvin06} propose an method automata to verify linearizability
of concurrent linked-list implementations, but only for the
execution of two fixed operations. Vechev et al. \cite{spin09} use SPIN
model checker to model and verify linearizability. Burckhardt et
al. \cite{pldi10} present an automatic linearizability checker Line-Up based
on the model checker CHESS.

To verify progress properties, Gotsman etal. \cite{Gotsman11} first propose
a generalisation of linearizability such that it can specify liveness.
Based on the definition, they reveal a connection between lock-free property
and a termination-sensitive contextual refinement, but they do not discuss other progress properties.
Liang etal. \cite{Feng,Feng14} study five progress properties and show that each
progress property is equivalent to a specific termination-sensitive
contextual refinement. But they need to define different refinement
relations for different progress properties.
In our work, divergence-sensitive branching bisimulation implies LTL$_{\_\bigcirc}$ equivalence and can preserve all
progress properties that are specified in LTL$_{\_\bigcirc}$. So there is no need to define different
relations for different progress properties.
Further, our recent work \cite{arXiv} extends the proposed bisimulation method such that it can verify
more complicated concurrent objects as in \cite{Feng13,Feng14}.


\paragraph*{Conclusion}
This paper proposes a novel and efficient method based on divergence-sensitive branching
bisimulation for verifying linearizability and lock-free property of concurrent objects.
Instead of using trace refinement to check linearizability,
our approach enables the use of branching bisimulation techniques
for checking both the linearizability and progress properties of concurrent objects.
Our experiment presents that (divergence-sensitive) branching bisimulation is an inherent equivalence relation
between the Treiber's lock-free stack (and the revised stack with hazard pointers)
and the corresponding concurrent specification.
Our method also reveals a new bug violating the lock-freedom.


\section*{Acknowledgment}
The first author would like to thank Xinyu Feng for the useful discussions on the area of
concurrent data structures.






\bibliographystyle{elsarticle-num}

\end{document}

%% file: macro.tex
\newtheorem{Def}{Definition}
\newtheorem{Expl}{Example}
\newtheorem{Thm}{Theorem}
\newtheorem{Lem}[Thm]{Lemma}
\newtheorem{fac}[Thm]{Fact}
\newtheorem{Cor}[Thm]{Corollary}

\def\squareforqed{\hbox{\rlap{$\sqcap$}$\sqcup$}}
\def\qed{\ifmmode\squareforqed\else{\unskip\nobreak\hfil
\penalty50\hskip1em\null\nobreak\hfil\squareforqed
\parfillskip=0pt\finalhyphendemerits=0\endgraf}\fi}

\newcounter{statement}
\def\stmnum{\hbox to .01pt{}\rlap{\rm \hskip -\displaywidth\thestatement.}}
\def\stm{\refstepcounter{statement}\topsep 2pt \trivlist \item[]\leavevmode
\hbox to\linewidth\bgroup $ \displaystyle \hskip\leftmargini}
\def\endstm{$\hfil \displaywidth\linewidth\stmnum\egroup \endtrivlist}

\newenvironment{Proof}{%
\begin{list}{}{\setlength{\topsep}{\jot}\setlength{\parsep}{\topsep}%
\addtolength{\parsep}{-0.3\parsep}\setlength{\leftmargin}{0pt}}%
\parindent 4ex
\item[]\setcounter{statement}{0}\textbf{Proof:}}{\end{list}}

\newtheorem{theorem}{Theorem}[section]
\newtheorem{lemma}[theorem]{Lemma}
\newtheorem{definition}[theorem]{Definition}
\newtheorem{notation}[theorem]{Notation}
\newcommand{\scarrow}[2]
      { \mathrel{\setbox0 \hbox{ ${\scriptstyle #1}$ }\displaystyle
            \mathop{\hbox to \wd0{$\Longrightarrow$}}\limits_{#1}^{#2}}\,\,\,
      }
\newcommand{\lscarrow}[2]
      { \mathrel{\setbox0 \hbox{ ${\scriptstyle #1}$ }\displaystyle
                 \mathop{\hbox to \wd0{$=\!=\!\Longrightarrow$}}\limits_{#1}^{#2}}
      }
\newcounter{LLN}
\newcommand{\beginit} {
                         \begin{list}{\hfill{\rm \arabic{LLN}.} }{\usecounter{LLN}
                         \setlength{\leftmargin}{\leftmarginii}
                         \setlength{\itemsep}{\smallskipamount}}}

\newbox\mystrutbox
\setbox\mystrutbox=\hbox{\vrule height10pt depth4pt width0pt}
\def\mystrut{\relax\ifmmode\copy\mystrutbox\else\unhcopy\mystrutbox\fi}
\newcommand{\lprf}[3]
       {\llap{{$#3$}\thinspace}{{\mystrut\displaystyle #1}
          \over
         {\mystrut\displaystyle #2}}}
\newcommand{\rprf}[3]
       {{{\mystrut\displaystyle #1}
          \over
         {\mystrut\displaystyle #2}}{\rlap{{$#3$}\thinspace}}}

\newcommand{\dotminus}{-\!\!\!\!^{\textstyle.}\!\!\!\!-}
\newcommand{\lit}[1]{{\rm [#1]}}
\newcommand{\henv}[1]{#1}
\newcommand{\setof}[2]{\{ #1 \: | \: #2 \}}
\newcommand{\bigsetof}[2]{\bigl\{ #1 \: \bigm| \: #2 \bigr\}}
\newcommand{\arrow}[1]{\stackrel{#1}{\longrightarrow}}
\newcommand{\sarrow}[1]{\stackrel{#1}{\Longrightarrow}}
\newcommand{\dobarrow}[1]{\stackrel{#1}{\Longrightarrow}}
\newcommand{\proc}{Pr}
\newcommand{\af}[1]{\forall{\rm F\;}#1}
\newcommand{\ef}[1]{\exists{\rm F\;}#1}
\newcommand{\ag}[1]{\forall{\rm G\;}#1}
\newcommand{\eg}[1]{\exists{\rm G\;}#1}
\newcommand{\varu}[3]{U^{#1}_{#2,#3}}
\newcommand{\varusfg}{\varu{s}{F}{G}}
\newcommand{\varuwfg}{\varu{w}{F}{G}}
\newcommand{\semu}[3]{{\cal U}^{#1}_{#2,#3}}
\newcommand{\semuwfg}{\semu{w}{F}{G}}
\newcommand{\semusfg}{\semu{s}{F}{G}}
\newcommand{\transu}[3]{{\cal T}^{#1}_{#2,#3}}
\newcommand{\transusfg}{\transu{s}{F}{G}}
\newcommand{\transuwfg}{\transu{w}{F}{G}}
\newcommand{\comp}[1]{{\cal C}(#1)}
\newcommand{\prc}[1]{{\cal P}(#1)}
\newcommand{\comppred}[1]{\prec_{#1}}
\newcommand{\bisim}[1]{\sim_{#1}}
\newcommand{\fat}[1]{\mbox{\bf #1}}
\newcommand{\tr}{\mbox{\rm tt}}
\newcommand{\fa}{\mbox{\rm ff}}
\newcommand{\act}{Act}
\newcommand{\impliess}{\: \Rightarrow \:}
\newcommand{\implied}{\Leftarrow}
\newcommand{\implieds}{\:\Leftarrow\:}
\newcommand{\iffs}{\:\Leftrightarrow\:}
\newcommand{\hviss}{\Leftrightarrow}
\newcommand{\sigent}{\sigma_{\entail{}}}
\newcommand{\ua}[2]{#1 \subseteq \sem{\D}#1 \cup #2}
\newcommand{\id}{\mbox{\rm Id}}
\newcommand{\powerset}[1]{{\Large \wp} (#1)}
\newcommand{\sem}[1]{\lbrack\!\lbrack #1 \rbrack\!\rbrack}
\newcommand{\abs}[1]{|\!| #1 |\!|}
\newcommand{\may}[1]{\langle #1 \rangle}
\newcommand{\wmay}[1]{\langle\!\langle #1 \rangle\!\rangle}
\newcommand{\wmust}[1]{\sem{#1}}
\newcommand{\until}[1]{[ #1 \rangle}
\newcommand{\smay}[1]{\langle\!\cdot #1 \cdot\!\rangle}
\newcommand{\must}[1]{[ #1 ]}
\newcommand{\smust}[1]{\lbrack\!\cdot #1 \cdot\!\rbrack}
\newcommand{\sat}[1]{\models_{#1}}
\newcommand{\mx}{{\rm max}}
\newcommand{\mn}{{\rm min}}
\newcommand{\satmn}{\sat{\mn}}
\newcommand{\satmx}{\sat{\mx}}
\newcommand{\sigmax}{\sigma_{\mx}}
\newcommand{\sigmin}{\sigma_{\mn}}
\newcommand{\entail}[1]{\vdash_{#1}}
\newcommand{\entmn}{\entail{\mn}}
\newcommand{\entmx}{\vdash}
\newcommand{\satt}[1]{|\!\!\!\equiv_{#1}}
\newcommand{\sattmx}{\satt{\mx}}
\newcommand{\sattmn}{\satt{\mn}}
\newcommand{\Dtu}{\D^{\tau}}
\newcommand{\Dtd}{\D_{\tau}}
\newcommand{\da}[2]{\semd #1 \cap #2 \subseteq #1}
\newcommand{\ass}[2]{#1 : #2}
\newcommand{\MM}[1]{{\cal M}_{#1}}
\newcommand{\D}{{\cal D}}
\newcommand{\mmid}{{\cal M}_{{\rm Id}}}
\newcommand{\og}{\wedge}
\newcommand{\eller}{\vee}
\newcommand{\semd}{\sem{\D}}
\newcommand{\M}{{\cal M}}
\newcommand{\infrule}[3]
           {\parbox{2cm}{ $$ {\frac {#1}{#2}}\hspace{.5cm}{#3} \hfill $$}}
\newcommand{\infrulegen}[4]
           {{#1}\hspace{.5cm}{\frac {#2}{#3}}\hspace{.5cm}{#4}}
\newcommand{\ent}{\entail{}}
\newcommand{\Gammah}{\widehat{\Gamma}}
\newcommand{\inv}[1]{\textsf{\rm Inv}(#1)}
\newcommand{\pos}[1]{\mbox{\rm Pos}(#1)}
\newcommand{\inva}{\inv{\may{a}\tr}}
\newcommand{\posa}{\pos{\must{a}\fa}}
\newcommand{\op}{{\cal O}}
\newcommand{\even}[1]{\mbox{\rm Even}(#1)}
\newcommand{\unic}{\bigr( \lbr\;|\;\hole\bsl p\bigr) \bsl\coin,\cof}
\newcommand{\live}{\mbox{\rm Live}}
\newcommand{\con}{\mbox{\rm Con}}
\newcommand{\com}{\mbox{\rm Com}}
\newcommand{\scom}{\mbox{{\cal Sc}}}
\newcommand{\dead}{\mbox{\rm Dead}}
\newcommand{\diver}{\mbox{\rm Div}}
\newcommand{\sUntil}[2]{\mbox{\rm Unt}^s(#1,#2)}
\newcommand{\wUntil}[2]{\mbox{\rm Unt}^w(#1,#2)}
\newcommand{\saf}[1]{\mbox{\rm Saf}(#1)}
\newcommand{\uni}{\mbox{\sl Uni}}
\newcommand{\staff}{\mbox{\sl Staff}}
\newcommand{\equip}{\mbox{\sl Equip}}
\newcommand{\acc}{\mbox{acc}}
\newcommand{\del}{\mbox{del}}
\newcommand{\wip}[2]{\mbox{\rm wip}(#1,#2)}
\newcommand{\sop}[2]{\mbox{\rm sop}(#1,#2)}
\newcommand{\cuni}{C_{\mbox{\rm uni}}}
\newcommand{\lbr}{\mbox{\rm lbr}}
\newcommand{\carrow}[2]{\raisebox{0.2ex}{$\,\,\,\,\,{#2\atop #1}
\!\!\!\!\!\!\!\!\arrow{}\,\,
$}}
\newcommand{\ccarrow}[2]{\raisebox{0.2ex}{$\,\,\,\,\,\,{#2\atop #1}
\!\!\!\!\!\!\!\!\!\arrow{}\!\!\!\!\!\!\!\!\arrow{}\,\,
$}}
\newcommand{\dccarrow}[2]{\raisebox{0.2ex}{$\,\,\,\,\,\,{#2\atop #1}
\!\!\!\!\!\!\!\!\!\arrow{}\!\!\!\!\!\!\!\!\arrow{}_\C\,\,
$}}
\newcommand{\cdarrow}[2]{\raisebox{0.2ex}{$\,\,\,\,\,\,{#2\atop #1}
\!\!\!\!\!\!\!\!\!\arrow{}_\C\,
$}}
\newcommand{\dcarrow}[2]{{\raisebox{-1.2ex}{
                         $\stackrel{#2}{\stackrel{\longrightarrow_\Diamond}
                          {\scriptstyle #1}}$  }}}
\newcommand{\rearrow}[2]{\raisebox{-1.2ex}{
                         $\stackrel{#1}{\stackrel{\longrightarrow}
                          {\scriptstyle #2}}$  }}
\newcommand{\bsl}{\backslash}
\newcommand{\hole}{\mbox{$[\;]$}}
\newcommand{\coin}{\mbox{coin}}
\newcommand{\cof}{\mbox{cof}}
\newcommand{\unicp}{\bigr( \cof .(\lbr + p.\lbr) \;|\;\hole\bsl p\bigr)
                                   \bsl \coin,\cof}
\newcommand{\unicpp}{\bigr( (\lbr + p.\lbr)\;|\;\hole\bsl p\bigr)
                             \bsl \coin,\cof}
\newcommand{\synen}[2]{#1\,\,{\bf{\sf with}}\,\,#2}
\newcommand{\emax}[2]{\mbox{{\bf {\sf max }}}#1\,\,{\bf{\sf with}}\,\,#2}
\newcommand{\emin}[2]{\mbox{{\bf {\sf min }}}#1\,\,{\bf{\sf with}}\,\,#2}
\newcommand{\lmax}{\mbox{{\sf max}\,}}
\newcommand{\lmin}{\mbox{{\sf min}\,}}
\newcommand{\Dsem}[1]{{\sf D} \lbrack\!\lbrack #1 \rbrack\!\rbrack}
\newcommand{\Delsem}[1]{{\sf \Delta} \lbrack\!\lbrack #1 \rbrack\!\rbrack}
\newcommand{\Asem}[1]{{\sf E} \lbrack\!\lbrack #1 \rbrack\!\rbrack}
\newcommand{\Lsem}[1]{{\sf L} \lbrack\!\lbrack #1 \rbrack\!\rbrack}
\newcommand{\Nsem}[1]{{\sf N} \lbrack\!\lbrack #1 \rbrack\!\rbrack}
\newcommand{\DMTSt}{\mbox{DMTS$^2$}\,}
\newcommand{\DMTSs}{\mbox{DMTS$^\ast$}\,}
\newcommand{\conf}[2]{\langle #1,#2\rangle}
\newcommand{\chop}[1]{\mbox{{\tt chop}$(#1)$}}
\newcommand{\tl}[1]{\mbox{{\tt tl}$(#1)$}}
\newcommand{\eemax}[1]{\mbox{{\bf {\sf max }}}#1}
\newcommand{\eemin}[1]{\mbox{{\bf {\sf min }}}#1}
\newcommand{\pre}{\mbox{\,{\tt pre}\,}}
\newcommand{\conft}[1]{\langle #1\rangle}
\newcommand{\incon}{IC}
\newcommand{\hml}{{\cal M}}
\newcommand{\depth}{depth}
\newcommand{\dsat}{{\,\mbox{{\tt sat}}}}
\newcommand{\branb}{{\approx_b}}
\newcommand{\V}{{\cal V}}
\newcommand{\Z}{{\cal Z}}
\newcommand{\ins}[3]{\vdash^?_{#3}#1\colon#2}
\newcommand{\insp}[3]{\vdash^{#3}#1\colon#2}
\newcommand{\insn}[3]{\not\vdash_{#3}#1\colon#2}
\newcommand{\wi}[2]{{\cal W}(#1,#2)}


\newcommand{\DEF}               {\stackrel{\rm def}{=}}
\newcommand{\invo}              {\textsf{inv}}
\newcommand{\res}               {\mathit{res}}
\newcommand{\lin}               {\textsf{lin}}
\newcommand{\Abs}               {\textsf{Abs}}
\newcommand{\client}            {\textsf{Client}}
\newcommand{\Sys}               {\textsf{Sys}}
\newcommand{\Systau}            {\textsf{Sys1}}
\newcommand{\Sysinv}            {\textsf{Sys2}}
\newcommand{\Sysres}            {\textsf{Sys3}}
\newcommand{\Sysobj}            {\textsf{Sys4}}
\newcommand{\emp}               {\varepsilon}
\newcommand{\Ex}                {\bf Ex}
\newcommand{\Eva}               {\bf Eva}
\newcommand{\true}              {\mathit{true}}
\newcommand{\false}             {\mathit{false}}

\newcommand{\StackOp}           {\mathit{StackOp}}
\newcommand{\LesOp}             {\mathit{LesOp}}
\newcommand{\TryStackOp}        {\mathit{TryStackOp}}
\newcommand{\TryCollision}      {\mathit{TryCollision}}
\newcommand\ignore[1]{}
\newcommand{\call}              {\texttt{call}}
\newcommand{\ret}               {\texttt{ret}}

\newcommand{\xhookrightarrow}[1]  {\stackrel{#1}{\hookrightarrow}}

%% file: licsmacro.tex
\newcommand{\necc}[2]{#1_{\Box}.#2}
\newcommand{\perm}[2]{#1_{\Diamond}.#2}
\newcommand{\rec}{\mbox{\bf rec}}
\newcommand{\barrow}[1]{\stackrel{#1}
                   {\longrightarrow_\Box}}
\newcommand{\cutarrow}[1]{\stackrel{#1}
                   {\longrightarrow_{\odot}}}
\newcommand{\zarrow}[1]{\stackrel{#1}
                   {\Longrightarrow_\Diamond}}
\newcommand{\marrow}[1]{\stackrel{#1}
                   {\longrightarrow_m}}
\newcommand{\darrow}[1]{\stackrel{#1}
                   {\longrightarrow_\Diamond}}
\newcommand{\tilarrow}[1]{
		   {\,\,\,\,\,\,{#1\atop{}}\!\!\!\!\!\!\!\raisebox{-0.2ex}
		{$\sim$}\!\!\!\rightarrow_\Diamond}\,}
\newcommand{\refin}{\lhd}
\newcommand{\rrefin}{\rhd}
\newcommand{\U}{{\cal U}}
\newcommand{\one}{\mbox{\bf 1}}
\newcommand{\nil}{nil}
\newcommand{\R}{{\cal R}}
\newcommand{\F}{{\cal F}}
\newcommand{\W}{{\cal W}}
\newcommand{\WC}{{\cal W}_c}
\newcommand{\E}{{\cal E}}
\newcommand{\semnil}{\mbox{\rm nil}}
\newcommand{\semperm}[1]{\lceil #1_{\Diamond}\rceil}
\newcommand{\semnecc}[1]{\lceil #1_{\Box}\rceil}
\newcommand{\altsem}[1]{\{\!\lbrack #1 \rbrack\!\}}
\newcommand{\finsyn}{{\cal FS}}
\newcommand{\ol}[1]{\overline{#1}}
\newcommand{\C}{{\cal C}}

%% file: cntxmacro.tex
\newcommand{\csystem}
         {(\langle C_n^m\rangle_{n,m} , A , 
           \langle \arrow{}_{n,m} \rangle_{n,m})}
\newcommand{\proj}[2]{\Pi_{#1}^{#2}}
\newcommand{\ident}[1]{\mbox{I}_{#1}}
\newcommand{\pref}[1]{{#1}.(\,\,)}
\newcommand{\zero}{{\bf 0}}
\renewcommand{\comp}{\mbox{{\rm C}}}
\renewcommand{\nil}{{\bf 0}}
\newcommand{\fdb}[1]{{#1}^{\dagger}}
\renewcommand{\F}{\mbox{{\rm F}}}
\renewcommand{\R}{{\cal R}}
\newcommand{\asscomp}{\mbox{{\rm A}}_{\circ}}
\newcommand{\assprod}{\mbox{{\rm A}}_{\times}}
\newcommand{\distprod}{\mbox{{\rm D}}_{\circ}^{\times}}
\newcommand{\distfdb}{\mbox{{\rm D}}_{\times}^{\dagger}}
\newcommand{\idaxiom}{\mbox{{\rm I}}}
\newcommand{\zerocomp}{\mbox{{\rm Z}}_{\circ}}
\newcommand{\zeroprod}{\mbox{{\rm Z}}_{\times}}
\newcommand{\zerofdb}{\mbox{{\rm Z}}_{\dagger}}
\newcommand{\zerozero}{\mbox{{\rm Z}}\mbox{{\rm Z}}}
\newcommand{\zeronull}{\mbox{{\rm Z}}_0}
\newcommand{\projdef}{\mbox{{\rm P}}_d}
\newcommand{\projprod}{\mbox{{\rm P}}_{\times}}
\newcommand{\fixaxiom}{\mbox{{\rm F}}}
\renewcommand{\zero}[2]{{\cal O}_{#1}^{#2}}
\renewcommand{\arraystretch}{1.3}
\newcommand{\letmax}[2]{\mbox{{\bf {\sf letmax }}}#1\,\,\mbox{{\bf
{\sf in }}}#2}
\newcommand{\letmix}[2]{\mbox{{\bf {\sf let }}}#1\,\,\mbox{{\bf
{\sf in }}}#2}
\newcommand{\letmin}[2]{\mbox{{\bf {\sf letmin }}}#1\,\,\mbox{{\bf
{\sf in }}}#2}
\newcommand{\form}{{\cal F}}
\newcommand{\decl}{{\cal D}}
\newcommand{\alt}{\,\,\,|\,\,\,}
\newcommand{\Fsem}[1]{{\sf
F} \lbrack\!\lbrack #1 \rbrack\!\rbrack}
\newcommand{\Dmax}[1]{{\sf
D} \lbrack\!\lbrack #1 \rbrack\!\rbrack}
\newcommand{\Dmin}[1]{{\sf D} \lbrack\!\lbrack #1 \rbrack\!\rbrack}
\renewcommand{\tr}{\mbox{{\sf tt}}}
\renewcommand{\fa}{\mbox{{\sf ff}}}
\newcommand{\func}{\rightarrow}
\newcommand{\env}{{\cal E}}
\renewcommand{\sat}{\models}
\newcommand{\w}{{\cal W}}
\newcommand{\vc}{V^{\C}}
\newcommand{\valid}{\models}
\newcommand{\wvalid}{\models_{\times}}
\newcommand{\LC}{{\cal L}}
\newcommand{\RC}{{\cal R}}
\newcommand{\uu}[2]{\mbox{$\bigcup_{#1} #2$}}
\newcommand{\nn}[2]{\mbox{$\bigcap_{#1} #2$}}
\newcommand{\fuu}[2]{\mbox{$\bigvee_{#1}#2$}}
\newcommand{\fnn}[2]{\mbox{$\bigwedge_{#1}#2$}}
\newcommand{\Dmaxsem}[2]{{\sf D}_\nu\sem{#1}#2}
\newcommand{\Dminsem}[2]{{\sf D}_\mu\sem{#1}#2}
\newcommand{\B}{{\cal B}}
\newcommand{\maxd}{{\sf max}}
\newcommand{\mind}{{\sf min}}